\documentclass{emulateapj}

\usepackage{natbib}
\usepackage{graphicx}
\usepackage{latexsym}
\usepackage{amsmath}
\usepackage{appendix}
\usepackage{longtable}

\slugcomment{Draft v5.0}

\shorttitle{FIR at high z}
\shortauthors{De Rossi et al. 2018}

\begin{document}

\title{The Far Infrared Emission of the First Massive Galaxies}

\author{Maria Emilia De Rossi\altaffilmark{1,2}, George H. Rieke\altaffilmark{3}, Irene Shivaei\altaffilmark{3}, Volker Bromm\altaffilmark{4}, \\ and Jianwei Lyu\altaffilmark{3}}

\altaffiltext{1}{Universidad de Buenos Aires, Facultad de Ciencias Exactas y Naturales y Ciclo B\'asico Com\'un. Buenos Aires, Argentina}
\altaffiltext{2}{CONICET-Universidad de Buenos Aires, Instituto de Astronom\'{\i}a y F\'{\i}sica del Espacio (IAFE). Buenos Aires, Argentina}
\altaffiltext{3}{Steward Observatory, Department of Astronomy, University of Arizona, 933 North Cherry Avenue, Tucson, AZ 85721}
\altaffiltext{4}{Department of Astronomy, The University of Texas, 2515 Speedway, Stop C1400, Austin, Texas 78712-1205}

\begin{abstract}

Massive Population~II galaxies undergoing the first phase of vigorous star formation after the initial Population~III stage should have high energy densities and silicate-rich interstellar dust. We have modeled the resulting far-infrared spectral energy distributions (SEDs), demonstrating that they are shifted substantially to bluer (``warmer'') wavelengths relative to the best fitting ones at $z \approx 3$, and with strong outputs in the $10 - 40 \mu$m range. When combined with a low level of emission by carbon dust, their SEDs match that of Haro~11, a local moderately-low-metallicity galaxy undergoing a very young and vigorous starburst that is likely to approximate the relevant conditions in young Population~II galaxies. We expect to see similar SEDs at high redshifts ($z \gtrsim 5$) given the youth of galaxies at this epoch. In fact, we find a progression with redshift in observed galaxy SEDs, from those resembling local ones at $2 \lesssim z < 4$ to a closer resemblance with Haro~11 at 5 $\lesssim z < 7$. In addition to the insight on conditions in high redshift galaxies, this result implies that estimates of the total infrared luminosities at $z\sim 6$ based on measurements near $\lambda\sim 1\,{\rm mm}$ can vary by factors of 2 $-$ 4, depending on the SED template used. Currently popular modified blackbodies or local templates can result in significant underestimates compared with the preferred template based on the SED of Haro~11.

\end{abstract}


\section{Introduction}
%

The far-infrared (FIR) spectral energy distributions (SEDs) of star-forming 
galaxies provide unique clues to the state of their interstellar medium (ISM), 
as well as to the process of star formation within them. For example, 
\citet{sauvage1992} and \citet{popescu2011} model the radiative transfer in star forming galaxies, 
showing how the basic behavior can be explained in terms of two underlying 
components: 1) warm and localized dust heated by very young stars and with 
an SED peaking in the 50 -- 100~$\mu $m range, and 2) cooler and diffuse 
dust heated by an older stellar population with an SED peaking in the 100 -- 
200~$\mu $m range. SED templates show that the relative role of the first 
component increases as the luminosity in young stars increases, with the 
warm component dominating in luminous and ultraluminous infrared galaxies (LIRGs and ULIRGs) \citep{rieke2009}. 
For the highest luminosity  {\it local} galaxies, the mid-infrared ($\lambda < $ 30~$\mu$m) emission is 
suppressed \citep[e.g.,][]{rieke2009} by optical depth effects arising from the extreme ISM 
densities in the star-forming regions \citep[e.g.,][]{dacunha2010}. At 
{\it moderate redshifts} ($1 \lesssim z < 3$), there is less of this 
suppression and the SEDs even of ULIRGs resemble those of local LIRGs 
\citep{symeonidis2009, symeonidis2013, rujopakarn2013, kirkpatrick2015}. At $z\sim 2$, the suppression due to high optical depth begins only at luminosities $\gtrsim 3 \times 10^{12}$ L$_\odot$, an order of magnitude more luminous than locally \citep{shipley2016}. This behavior seems to be connected with the greater extent of the FIR sources in the very luminous infrared galaxies 
at these redshifts, compared with those in local galaxies of similar 
luminosity \citep{rujopakarn2011, shipley2016}.  In addition, \citet{schreiber2018} demonstrate a reduction in the peak wavelength 
(i.e., an increase in the equivalent temperature) of the far-infrared SEDs of star forming galaxies with redshift from $z \lesssim 1$ to $z \simeq 3.5$.

However, the behavior of star forming galaxy far-infrared SEDs at $z \gtrsim 3$ is not well characterized and 
thus cannot yet be used for similar insights about them.  In general, relatively little information is available to model the 
SEDs, so the behavior has been described in simplified ways.
\citet{wang2007} fitted the FIR SEDs (assumed to be powered by star formation) for high redshift quasars with a 
modified blackbody of temperature 47\,K and emissivity index $\beta$ = 1.6. 
These parameters have been adopted by many subsequent works \citep[e.g.,][]{leipski2014, willott2017}. 
\citet{bowler2018} used a similar approach but with a fiducial temperature of 40\,K. 
\citet{lyu2016} found that template fits to the SEDs of high redshift quasars were improved dramatically by using the relatively 
warm SED of the low metallicity, extreme star-forming galaxy Haro 11, in 
place of modified blackbodies or the colder templates that apply at redshifts up to $2-3$. \citet{faisst2017} 
used indirect arguments to conclude that the FIR SEDs 
of very high redshift galaxies in general are ``hotter'' than those of local 
galaxies. \citet{yuan2015} used color-color diagrams in the SPIRE bands (250, 350, and 500 $\mu$) to
show that by $z = 5 - 6$ none of the templates they considered \citep{chary2001, dale2002, elbaz2011, magdis2012, berta2013, ciesla2014} 
were consistent with the data.

A trend toward warmer SEDs would arise from the increasing 
temperature of the cosmic microwave background (CMB) with increasing 
redshift \citep{dacunha2013}. However, given the high temperatures we find for the SEDs ($>$ 40 K at $z \approx 6$), 
the effect is not large enough to 
explain the observed behavior; see fig.~5 of \citet{dacunha2013}.  As an example, we have applied equation (12) 
of \citet{dacunha2013} to parameters  for the coldest dust  contributing to the 
peak of the SED (T = 44 K, $\beta = 1.2$) that would maximize the CMB contribution for Haro 11 
(suggested as a suitable high redshift template by \citet{lyu2016}). We find that 
the CMB would raise the dust temperature at z = 6 by $<$ 1\%.
Any perceptible effect of the CMB would be modest and confined to the mm-wave regime \citep{dacunha2013}. 
\citet{strandet2016} also find that the CMB has negligible influence on far infrared SEDs at $z < 10$. 
Therefore, the possible trend must arise from heating of dust by young stars 
within the galaxies and be associated with differences in the stellar populations 
and/or the ISM at high redshift compared with local galaxies.

The general trend toward higher SED temperature with redshift indicates an increase 
from typical temperatures of $\sim$ 24K at z = 0 to $\sim$ 40K at z = 3.5 $-$ 4 and would predict T $\sim$ 
51 K at z = 6 \citep{schreiber2018}. However, fits to the SEDs of high redshift quasar host galaxies indicate that 
in addition there is a much larger contribution in the rest 10 $-$ 40 $\mu$m 
range from very warm dust than would be predicted just by shifting local templates toward higher temperature \citep{lyu2016}. 
 It is premature to try to model the far infrared SED behavior accurately {\it ab initio}; instead we focus on 
identifying a galaxy with a SED that can be used as a template for the high-redshift SEDs. To do so, we first discuss 
the available theoretical constraints. 
Section 2 of this paper shows that the very warm dust can be the result of the silicate-rich dust composition  
expected for galaxies $\lesssim$ 400 Myr in age, particularly at the high UV radiation density of very high redshift galaxies. 
However, there is a large range of possible optical properties and resulting SEDs from such dust. 
 It is not possible to sort out the ``best'' model for the silicate rich dust of the eight we consider, so we need to appeal to a real galaxy with conditions as close as possible to those expected for the high redshift galaxies. In Section 3 we show that conditions in Haro 11 match reasonably well those in luminous infrared galaxies at z $\sim$ 6, including having an environment conducive to its dust being silicate-rich. However, 
all of  the silicate-rich compositions predict SEDs that are weak in the far infrared (at $\sim$ 100 $\mu$m) compared 
with the SED of Haro 11. We show that addition of a small amount of 
amorphous carbon dust to the silicate-rich dust in the theoretical models removes this discrepancy. 
The addition of this carbon is consistent with some models of star formation in the early Universe, and is also required by the 
detection of the 158 $\mu$m carbon line in a number of z $\sim$ 6 galaxies \citep[e.g.,][]{pentericci2016, bradac2017, decarli2017, smit2018}. 
We therefore adopt the SED of Haro 11 as an empirical template for comparison with the measurements of high redshift galaxies. 
This comparison is carried out in Section 4, where we show that this template is a better fit to the data at z $>$ 5 
than any of the other proposed possibilities. Understanding FIR SEDs at 
high redshift is important to calculate total luminosities and to calibrate determinations of star formation 
rates (SFRs). Section 5 discusses how the selection of a Haro-11-based template affects luminosities from the existing database of 
Herschel measurements. Additional future determinations will largely utilize continuum measurements with the 
Atacama Large Millimeter/submillimeter Array (ALMA). Most of 
the observations with ALMA are made at about 1 mm, so we also discuss the template-dependence of the deduced luminosity 
from such data. The major results of the paper are summarized in Section 6.   

\section{Modeling the Far Infrared SEDs of High Redshift Galaxies}

\subsection{General Behavior of Population II Galaxies}

Our SED models in the FIR are based on those developed for primeval galaxies\footnote{We use the following terminology: (1) Population~III (Pop~III) stellar systems contain metal-free stars 
that begin the enrichment of their ISMs through type II supernovae; and (2) Population~II (Pop~II) galaxies are the descendants of Pop~III
stellar systems with metal-poor stars that continue to enrich the ISM through an increasing variety of processes as they evolve. An extreme case within the latter category are ``one-shot'' systems where the metal enrichment reflects pure Pop~III nucleosynthesis \citep{frebel2012}. However, the exact duration 
of the intense star forming episode in Pop~II galaxies is not well-constrained; a plausible range is 0.1 - 0.5 Gyr \citep[e.g.,][]{kriek2016}. In this paper, we focus on the period up to $\sim$ 400 Myr past the initial formation of stars, representing the primeval galaxy stage, during which stars 
less massive than 3.5 M$_\odot$ have not ejected significant amounts of carbon-enriched material into the ISM.} by \citet{derossi2017}, hereafter DB17.  According to their findings, the FIR SEDs of these galaxies are likely to be significantly warmer than typical FIR SEDs of galaxies in the $0<z<3$ range. 
Our goal is to explore this transition between $z = 3$ and $z \sim 10$. 
Such behavior likely arises in large part because of the high energy density in young Pop II galaxies combined with the characteristics of the silicate-rich dust expected in systems that are dominated by Pop~III enrichment. Silicates have relatively poor emission efficiency at wavelengths short of about 8~$\mu$m and longer than about 60~$\mu$m \citep{koike2003}, accentuating the tendency from the high energy density for relatively high temperatures for the dust.  {\it The transition in FIR SEDs may therefore depend not on the transition from Pop III to Pop II in itself, or on the duration of the Pop II burst of star formation, but directly on the transition away from silicate-rich dust.} We will adapt the model of DB17 to conditions in the galaxies being detected in the far infrared 
at redshifts of $z \sim 6$ to explore this possibility.

The composition of Pop III dust is not well known \citep[e.g.,][]{schneider2016,jaacks2018}. As a specific example, \citet{ji2014} considered silicon-based dust models taken from
\citet{cherchneff2010}, and DB17 calculated the FIR SEDs for these different models.
It is worth mentioning that, for estimating dust chemical composition, \citet{cherchneff2010} assumed non-equilibrium chemical
kinetics for dust formation, while most steady-state models would predict a more significant carbon composition.
The suppression of carbon in the former case is related to the hypothesis that carbon-rich regions in the supernova
ejecta are microscopically mixed with helium ions. As in DB17, we follow these assumptions to probe how 
silicate-rich dust from Pop III/II stars can affect the infrared SED; we will return to the issue of carbon dust in Section 3. 

By analyzing the different dust chemical models implemented in \citet{ ji2014}, DB17 found that,  for galaxies with similar properties, the FIR intensities and general shapes of the predicted SEDs are similar, but the detailed spectral features show differences. 
As a result, the theoretical results can only be taken as representative; alternative mixtures of 
silicate-rich dust would alter the predicted SEDs, although the overall shape will remain controlled by 
the poor emission efficiency of this type of mineral outside the 8~$\mu$m to  60~$\mu$m range. Improving the correspondence 
between the predicted and an observed SED would require adjustments in the adopted silicate material that would have to be 
{\it ad hoc} in nature, given the lack of any other meaningful observational constraints. 

Rather than attempting such adjustments, in the following section we will propose the galaxy Haro 11 as a kind of analog computer that can utilize empirically determined dust optical properties. As we discuss below, Haro 11 has an exceptionally high rate of star formation in a low-mass 
and (at least prior to the ongoing starburst) low metallicity galaxy.  
These conditions are conducive to its interstellar dust being dominated by the outputs of very young and massive stars, as is the case for the high redshift galaxies. The energy density in the dominant star forming region in this galaxy is also similar to that in the high redshift galaxies. We will show that the FIR SED of this galaxy generally resembles that predicted by the theoretical model.  \citet{lyu2016} have also suggested 
that the Haro 11 SED is a good match to the FIR SEDs of quasar host galaxies at z $\sim$ 6. We will therefore adopt its SED as a proxy  for the theoretical one and 
use it in the remainder of the paper to compare with far infrared observations of high redshift galaxies.

\subsection{Conditions in the Modeled Galaxies }

\subsubsection{Dust Composition}

Gas that is only enriched by Pop~III nucleosynthesis will have a unique chemical signature that will be erased as the galaxy 
evolves to Pop II \citep{ji2015}, most notably when stars with masses $<$ 3.5 M$_\odot$ enter the asymptotic giant branch (AGB) phase and yield large amounts of carbon.
The calculations of DB17 hence 
assumed a silicate-rich dust composition \citep[e.g.,][]{ji2014, marassi2015} produced by the type II supernovae  that 
dominate the dust production in Pop III stellar systems \citep{dicriscienzo2014}. The type II supernovae
are also likely to be a significant source of dust in very young 
galaxies that evolve from Pop III \citep[e.g.,][]{asano2011, ji2015}.  AGB stars will be another potential 
dust source in Pop II. \citet{ventura2012a,ventura2012b} find that low-metallicity 
massive AGB stars produce silicate-rich dust, while less massive ones produce 
substantial carbon, with the transition mass at 3.5 M$_{\odot}$. 
There may also be an additional less-abundant component of corundum/alumina 
\citep{ventura2014,dellagli2014} from massive AGB stars. These 
theoretical predictions are confirmed observationally, e.g. by \citet{ventura2015}. A star of 3.5 M$_{\odot}$ has a main sequence 
lifetime of $\sim$ 400 Myr, suggesting that the silicate-rich dust 
composition should dominate in galaxies younger than this age. This transition 
is likely to be more critical to the FIR output of a galaxy than the end of the Pop~III stage of evolution.

\citet{michalowski2011} and \citet{michalowski2015} find that the dust produced directly in AGB stars 
and type II supernovae may be insufficient to explain the amounts of dust 
found in high redshift galaxies, and that much of the dust must form in the 
ISM. However, this process is likely to be delayed past the timescale of interest for this paper - i.e., 
before significant carbon is injected into the ISM \citep{asano2011}. 

\subsubsection{Sizes and SFR Surface Densities}

To be detectable at current limits, individual high-redshift galaxies must have very high FIR luminosities. The median for those in our study at $5 \le z < 7$ is  $5.3 \times 10^{13}$ L$_\odot$ (Section 3.2), which becomes $\gtrsim 10^{13}$ L$_\odot$ even allowing for the median level of lensing amplification \citep{spilker2016}. The latter luminosity indicates a SFR $\gtrsim$ 1000 M$_\odot$ yr$^{-1}$, assuming a typical local initial mass function. 
The star forming regions in these galaxies are also 
compact \citep{spilker2016}. Specific examples are sizes of 2.5$\times$1.1 kpc at $\lambda_{\rm rest} \sim 160 \mu$m for AzTEC3 with a 
SFR of 1100 M$_{\odot}$\,yr$^{-1}$ \citep{riechers2014}, 1.8$\times$1.3 kpc 
and 2.1$\times$0.9 kpc at $\lambda_{\rm rest} \sim 130 \mu$m for the two components of ADFS-27 with a combined SFR of 
2400 M$_{\odot}$\,yr$^{-1}$ \citep{riechers2017}, and 1.22 kpc and 4500 
M$_{\odot}$\,yr$^{-1}$ also at $\lambda_{\rm rest} \sim 130 \mu$m for SPT 0346-52 \citep{ma2016}. The quasar host galaxies must also be very compact, given the difficulty in imaging them from underneath the glare of the quasar \citep[e.g.,][]{hutchings2003, hutchings2005, mcleod2009, targett2012, metchley2012}. In fact, the dust continuum and [C II] emission of quasar host galaxies at $z\gtrsim 6$ typically have sizes of $\sim$1--3 kpc (diameter) as revealed by ALMA \citep[e.g.,][]{wang2013, decarli2018}. Typical SFRs for these sources are 1000 - 5000 M$_\odot$ yr$^{-1}$ \citep{lyu2016}, presumably unlensed. 
These values suggest SFR surface densities of 300 -- 2000 M$_{\odot}$/(yr 
kpc$^{2})$. Under a broad range of conditions, the high luminosity 
density that results will produce a warmer FIR SED than would be 
found in lower luminosity density sources, such as normal nearby star 
forming galaxies \citep[e.g.,][]{chakrabarti2005, hayward2011}.

\subsubsection{Metallicity}

Comparison of the recent rest-frame UV spectroscopic observations and 
optical broad-band photometry with photoionization models indicates 
metallicities of $Z=$0.0016 $-$ 0.0026 for a small sample of galaxies at 
$z>6$ \citep{stark2017, mainali2017}, which is equivalent 
to 0.10 $-$ 0.17 Z$_{\odot}$ (assuming Z$_{\odot}=$ 
0.015, \citealt{bressan2012}). 
Moreover, \citet{hashimoto2017} find that low 
metallicities ($<$ 0.02 Z$_\odot$) and young stellar ages 
($<$ 100 Myr) are a possible cause of the large Lyman-$\alpha$ equivalent 
widths of Lyman-$\alpha$ emitters at $z\sim 3-5$, which are 
representative of galaxies in the early stages of their evolution. However, 
these large equivalent widths can also be explained by anisotropic radiative 
transfer effects, fluorescence by hidden AGN or QSO activity, and/or 
gravitational cooling. Recently, comparison of hydrodynamical simulations 
with ALMA observations of CR7, a luminous Ly-$\alpha$ emitter at 
$z=6.6$, suggests metallicities of $Z=$0.1 $-$ 0.2 Z$_{\odot}$ for the star-forming clumps in this galaxy \citep{matthee2017}. As a 
summary, based on the current observations, the metallicity of massive 
star-forming galaxies at 5 $\lesssim z \lesssim$ 7 is estimated to be 
$\sim$ 0.1 $- $ 0.2 Z$_{\odot}$. Lower metallicities are, of 
course, expected for lower-mass galaxies and at higher redshifts \citep{bromm2011}. Future 
observations, e.g. with the {\it James Webb Space Telescope (JWST)}, are required to constrain the gas-phase 
metallicity of these high redshift galaxies better.

\subsection{Modeling the Galaxy Far Infrared SEDs}

\begin{figure}
\epsscale{1.150}
\plotone{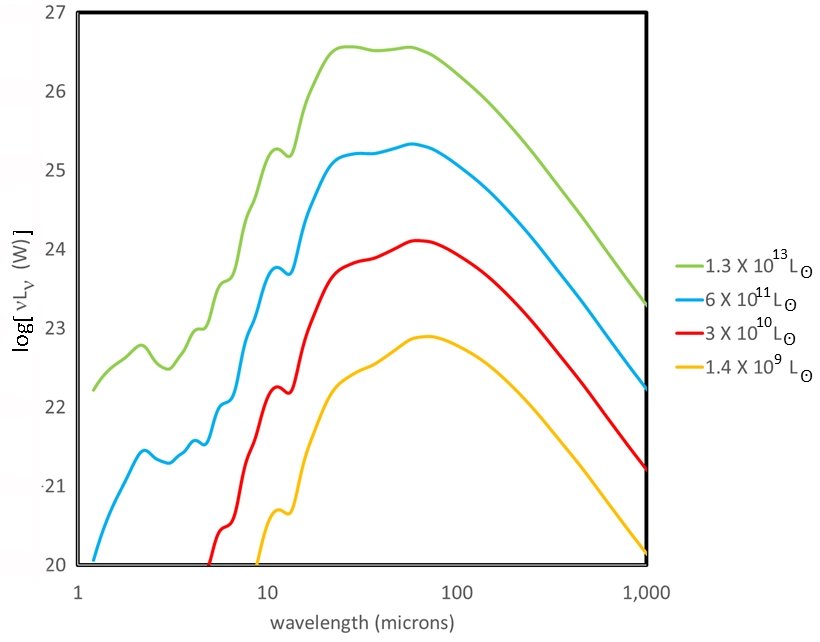}
\caption{Predicted far infrared SEDs for the modeled galaxies, as a function of their total infrared (5 -- 1000~$\mu$m) luminosities. The models are based on those discussed by DB17 with
 stellar mass fraction ($f_*$) of $M_*/(M_* + M_{\rm gas}) = 0.01$, modified to set the gas-phase metallicity to 0.33 Z$_\odot$, and dust-to-metal ratio to 0.02. The detectable galaxies at $z > 5$ have luminosities $\gtrsim 10^{13}$ L$_\odot$, similar to the most luminous case illustrated. For the sake of clarity, only SEDs from the UM-D-20 model are shown.
\label{fig1}}
\end{figure}

The calculations of FIR SEDs by DB17 were for
primeval galaxies dominated by Pop~II stars, and assumed 
extremely low metallicities and compact star forming complexes. 
Figure~\ref{fig1} shows the result of further developing their model 
in circumstances more applicable to 
the very luminous galaxies detectable at very high redshift, as described in Section 2.1.  To increase 
the far infrared outputs by $2-3$ orders of magnitude to match the high luminosities 
of the galaxies, we (1) 
increased the virial masses (M$_{\mathrm{vir}})$ of the modeled sources, (2) 
increased the gas metallicity to 1/3 that of the sun, and (3) assigned a dust-to-metal ratio of 0.02. 
 In particular, the SEDs associated with different luminosities in Figure~\ref{fig1} correspond to sources with halo mass M$_{\mathrm{vir}} = 10^{10}, 10^{11}, 10^{12}$ and 
$10^{13} \ {\rm M}_{\sun}$.
As in DB17, the  stellar mass fraction is set to $ f_* = M_*/(M_* + M_{\rm gas}) = 0.01$.

We evaluated eight of the sets of dust optical constants from \citet{ji2014} to sample the resulting variations in the SEDs; to suppress sharp features we smoothed the optical constants with a logarithmic  full width at half maximum  of 0.075 (i.e., a factor of 1.19). Specifically for Figure \ref{fig1} we used the dust composition UM-D-20 (see table~1 in \citealt{ji2014}), which is silicate-rich with some alumina, and the ``standard'' size distribution from \citet{ji2014}. As we will show, the implementation of other silicon-based dust chemistries generally predicts similar shapes of the SEDs, but the detailed features exhibit some variations. The use of other size distributions for the dust grains can lead to moderate variations in the total luminosities, with the exact value depending on the dust chemical composition. As discussed in \citet{ji2014}, a variety of dust compositions and size distributions are possible; our calculations are  illustrative of the far infrared SEDs that might be expected from very young galaxies,  but the detailed spectral features are a product of the specific grain composition assumed and should not be taken to be unique.  For further details regarding these choices, we refer the reader to DB17.
 
The models assume that the galaxy size is given by the virial radius. This has the effect that within a given family of models, such as those 
displayed in Fig.~\ref{fig1}, the mass-density is similar, independent of luminosity. The luminosity density exhibits a modest dependence, increasing by 
an order of magnitude over the range of luminosities in Fig.~\ref{fig1}. 
As shown by DB17, moderate changes in the gas-phase and dust density profiles (at a fixed mass) would not
generate significant changes in the SEDs, but an increase of the dust-to-metal ratio or gas-phase metallicity
drives an almost proportional increase of emission by dust without substantially changing the shape of the SED.

\begin{figure}
\epsscale{1.150}
\plotone{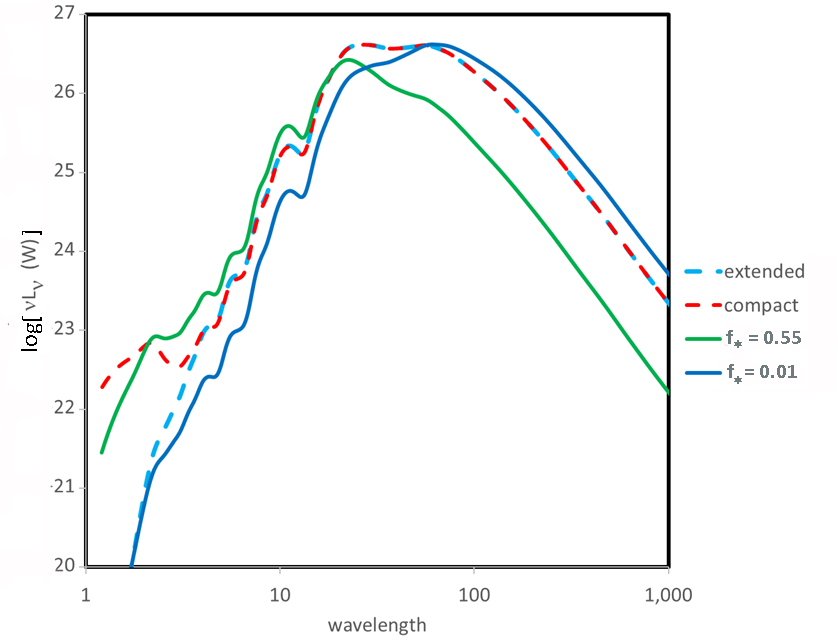}
\caption{Examples of changes in the theoretical SED with changes in the distribution of heating sources and in the luminosity density, expressed in terms of the  stellar mass fraction ($f_*$). The model for compact heating assumes a single, central luminosity source, while the extended heating arises from sources randomly distributed within a 100 pc radius. The luminosity density increases by an order of magnitude between $\bf f_*$ = 0.01 and $\bf f_*$ = 0.55.
\label{fig2}}
\end{figure}

Figure \ref{fig2} demonstrates how different assumptions affect the SEDs. All four SEDs in this figure 
are for a luminosity of $\sim$ $4 \times 10^{10}$ L$_\odot$. In particular, we analyze the effects of extending the heating source by distributing individual stellar groups over a radius of 100~pc.
The cases for an extended star cluster vs. a single central source merge for wavelengths $>$ 5~$\mu$m, so the distribution of heating sources 
to first order has relatively little effect on the predicted mid- and far-infrared SED. The increase in stellar mass fraction 
(sometimes called star formation efficiency) is accompanied by a decrease 
in the virial radius by a factor of 0.45, so there is an increase by an order of magnitude in the luminosity density that manifests itself as a modestly ``bluer'' SED, shifted to shorter wavelengths by a factor of about two. Thus, the far infrared SED changes with increasing luminosity density, but not dramatically. These results show that the emission efficiency of the silicate-rich dust has a strong influence on the FIR SED, making it relatively insensitive to changes in 
the architecture of the galaxy.

\section{Comparison with SED of Haro 11}

\subsection{Conditions in Haro 11 compared with those in high redshift galaxies}

\begin{figure}
\epsscale{1.150}
\plotone{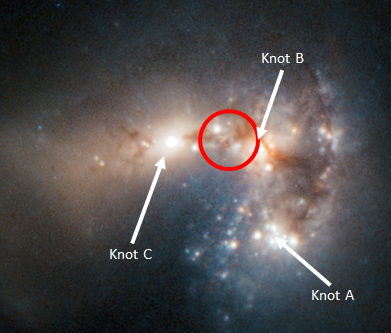}
\caption{Image of Haro 11 from \citet{adamo2010}. The red circle (2$''$ in diameter, or 700 pc) shows the location of the 
24 $\mu$m source and its diameter is roughly the upper limit on the size that can produce the majority of its signal, as 
implied by the diffraction-limited {\it Spitzer} image \citep{lyu2016}. 
Credit: ESA/Hubble/ESO and NASA.
\label{fig3a}}
\end{figure}

We now consider why Haro 11 is a useful analog to luminous, high redshift galaxies. 
The visible image of Haro 11 is complex with three bright knots (A, B, and 
C) within a fainter surrounding, the general appearance of a complex merging 
system, and with a rich population of very young super star clusters (e.g., Adamo 
et al. 2010). However, the very luminous infrared emission is compact and 
centered on Knot B \citep[e.g.,][]{lyu2016, chu2017}); see Figure~\ref{fig3a}, and indicates a SFR of about 30 M$_{\odot}$~yr$^{-1}$ from this 
knot \citep{lyu2016}. This knot has a higher metallicity than the rest of the galaxy \citep{james2013}, consistent 
with it being a site dominated by this ongoing starburst. Knot B has a radius of 
about 140 pc and a mass of about $2 \times 10^9$ M$_{\odot}$, determined dynamically \citep{ostlin2015}.
Its specific star formation rate is thus $\sim$ 15 Gyr$^{-1}$ and its SFR surface density is 
$\sim$ 500 M$_\odot$ yr$^{-1}$ kpc$^{-2}$; both values are typical of luminous high redshift galaxies. 

\begin{figure}
\epsscale{1.150}
\plotone{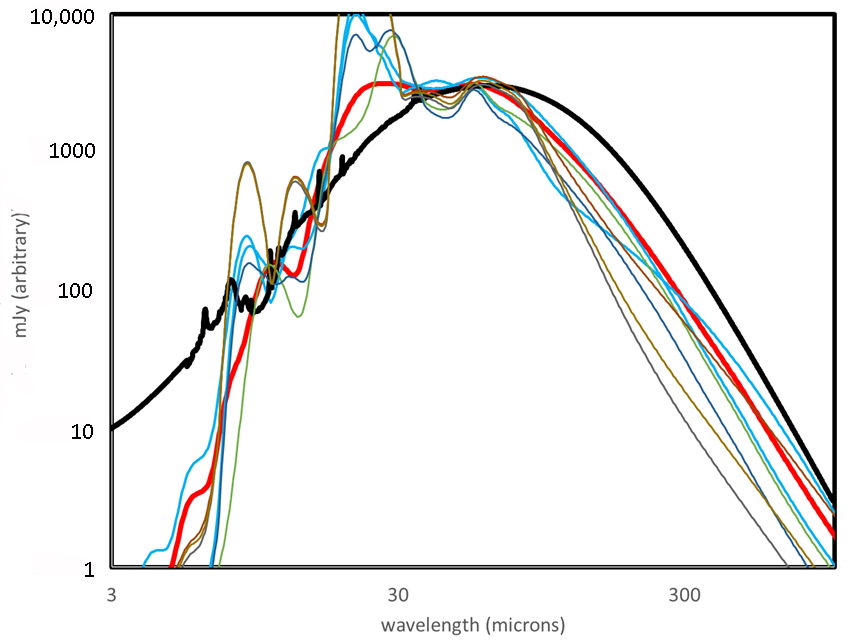}
\caption{Comparison of the FIR SED of the model from Fig.~1 and 2 for a galaxy of  $1.3 \times 10^{13}$ L$_\odot$ (thick red line) with that of Haro 11 (black). Seven alternative models using different possible optical constants (U-M-20, U-M-170, UM-D-170, 
M-ND-20, M-ND-170, UM-ND-20, UM-ND-170; see \citealt{ji2014}, for details) are also shown in finer lines.
\label{fig3b}}
\end{figure}

The stellar population powering the far infrared emission 
appears to be very young, with a typical age of only 3.5 Myr \citep{adamo2010}. A rough estimate of the total mass of very young stars is 3$\times$10$^8$ M$_{\odot}$, obtained by assuming 
the current level of SFR for 10$^{7}$ years.  A substantial fraction of 
the mass in Haro 11 arises from cold gas, which from its radial velocity 
resides primarily in Knot B \citep{cormier2014}. Indeed, if the gas mass is estimated from the far infrared emission, it accounts for most of the dynamical mass of $2 \times 10^9$ M$_{\odot}$ \citep{ostlin2015}.  Hence, the newly formed stars 
could be responsible for a very substantial fraction of the total {\it stellar} 
mass in Knot B; it is a region that is potentially young-star-dominated and 
hence a candidate to have silicate-rich dust. 

The oxygen abundance in Knot B is cited nominally as 12 $+$ log(O/H) $= 8.33 \pm 
0.01$, where the quoted error does not allow for systematic errors \citep{guseva2012}, or $8.25 \pm 0.15$ \citep{james2013}, values 
that are $\sim$ 36 -- 44{\%} of solar\footnote{taking the solar 12 $+$ log(O/H) 
$=$ 8.69}. This abundance is a factor of about three greater than our estimate for the high redshift ($z \sim 6$) 
galaxies.  The aromatic bands are very weak 
\citep{wu2006}, consistent with a 
carbon-poor ISM. However, other explanations are possible since the excitation and destruction of the 
aromatic bands is a complex and not well-understood process. 

In summary, the analogy of expected conditions in Haro 11 with those in luminous infrared 
galaxies at  $z\sim 6$  is reasonably good. 

\subsection{Comparison of Haro 11 SED with theoretical models}

In Figure~\ref{fig3b}, we compare the SEDs from our theoretical models with that 
for Haro 11 (from \citealt{lyu2016}, and references therein). As discussed previously, the model is conceptual - the lack of detailed information about the optical constants for the dust in silicate-rich galaxies means that the details of the SED are not significant. Nonetheless, some important trends are evident. First, the models do reflect the warmer temperature of Haro 11 reasonably well, and they {\it all} show the strong emission in the 10 $-$ 40 $\mu$m range, characteristic 
of the very warm dust that is also seen in the high redshift galaxies. The detailed behavior in this spectral range differs substantially from one set 
of optical constants to another. Even with much better constraints on the optical constants, a broad variety of behavior is also seen in spectra of local low metallicity galaxies where the silicate bands appear anywhere from strong emission 
to absorption \citep{remyruyer2015}.  For these local galaxies, despite the complete sets of observations, modeling the silicate bands in a general way is still 
challenging; \citet{remyruyer2015} had to introduce an additional spectral feature in the form of a warm modified blackbody to improve the fits 
to their observations. The selection of the appropriate optical constants among the possibilities for Pop III/II galaxies and detailed modeling of their spectra 
such as that for local galaxies by \citet{remyruyer2015} is beyond the scope of this paper.  

\begin{figure}
\epsscale{1.150}
\plotone{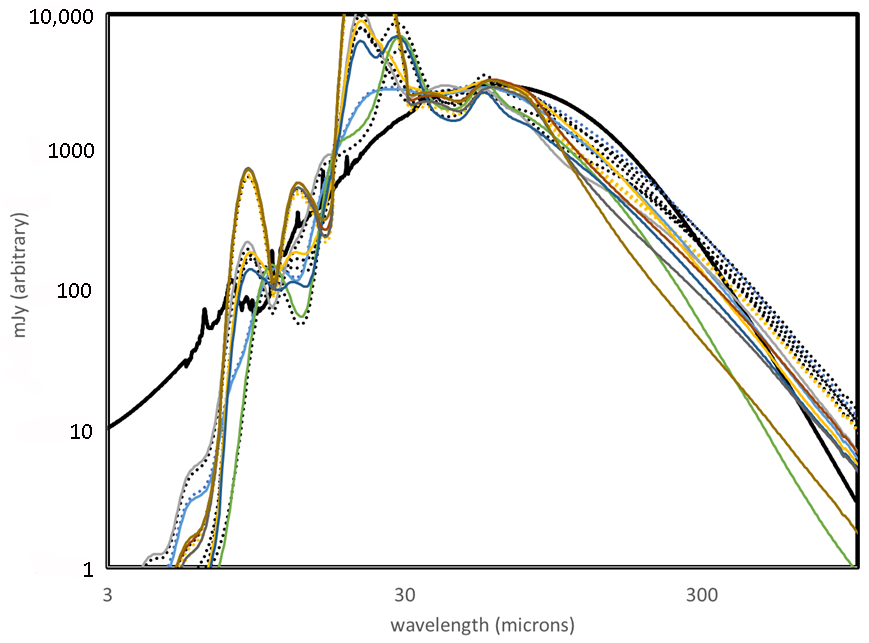}
\caption{The effect of adding emission from amorphous carbon dust to the eight types of silicate rich dust from \citet{ji2014}. The 
solid lines show the results if 10\% of the luminosity is from the carbon dust and the dotted lines show the results for 20\% coming from the carbon. The thick black line is the SED of Haro 11. 
\label{fig3c}}
\end{figure}

However, the deficiency in output at wavelengths $>$ 100 $\mu$m relative to the behavior of Haro 11, present in all models, needs to be addressed. There is 
ongoing discussion of whether there is a modest amount of carbon dust in the earliest galaxies \citep[e.g.,][]{nozawa2013, marassi2015}.  In any case, 
the detection of the [CII] 158 $\mu$m line in galaxies at $z > 5$ \citep[e.g.,][]{pentericci2016, bradac2017, decarli2017, smit2018} demonstrates that there 
is some level of this element in typical ISMs at this redshift, indicating also the presence of carbon dust. We therefore computed the infrared 
SED expected for amorphous carbon, using optical constants from Karl Misselt (private communication), based on those of \citet{zubko1996}. It is not 
possible to determine the abundance of carbon in these galaxies accurately from the 158 $\mu$m line detections \citep{lagache2018}. In addition, the 
emission efficiency of carbon depends strongly on its form (e.g., amorphous or graphite, see \citet{remyruyer2015}), so we do not show 
results as a function of relative densities of carbon and silicate dust. Instead, in Fig.~\ref{fig3c} we show the eight silicate models considered in 
Fig.\ref{fig3b}, but with 10\% and 20\% of the total luminosity contributed by carbon. It appears that the addition of a small amount of carbon dust can 
substantially improve the correspondence of the silicate-rich templates to the SED of Haro 11. 

Given the correspondence of the SED of Haro 11 to the model predictions,  we adopt the Haro 11 SED as a proxy template to study the observed 
behavior of FIR SEDs with redshift. We provide the Haro 11 SED in electronic form in Appendix A.

\section{Empirical Constraints on the High Redshift Far Infrared SEDs}

The preceding section has demonstrated the theoretical expectation that the SEDs will be relatively ``warm'' and broad for galaxies where: (1) a very recent episode ($<$ 400 Myr old) of star formation dominates the composition of the interstellar dust; and (2) the galaxy luminosity and size are appropriate for young galaxies at high redshift. We now compare far infrared and sub-mm measurements of galaxies at  $2\le z <7$ with this prediction. We begin with the range $2 \le z <4$  (median infrared luminosity of $6 \times 10^{12}$ L$_\odot$ for $z = 2 - 3$ and $1.8 \times 10^{13}$ L$_\odot$ for $z = 3 - 4$), 
demonstrating that the appropriate template derived from local luminous infrared galaxies provides a good fit to all the available far infrared/sub-mm observations. We next consider the range $5 \le z <7$  (median infrared luminosity $5.3 \times 10^{13}$ L$_\odot$), finding that the Haro 11-based template is the better fit. Finally, we consider the  $4\le z <5$  range (median infrared luminosity $3.2 \times 10^{13}$ L$_\odot$), finding that it shows hints of a transition from the lower redshift template to the Haro 11 one, i.e., it provides a bridge between the former cases. The quoted median luminosities are apparent, not allowing for lensing; it is expected that many of these sources are in fact lensed \citep{strandet2016}, with typical amplifications of $\sim$ 6 \citep{spilker2016}. Nonetheless, the changes we see in the far infrared SED are very different from the changes in local templates with increasing luminosity where, for example, the far infrared peak narrows  \citep{rieke2009}, so we believe that the unavoidable increase in detectable apparent luminosity with redshift cannot account for the change in SEDs.

\subsection{Far Infrared SEDs at $z\sim 3$}

At redshifts of 1 -- 3, \citet{rujopakarn2013} found that star formation 
rates estimated from rest-frame 24~$\mu $m photometry on the basis of templates from 
\citet{rieke2009} for log($L$(TIR)) $=$ 11.25 $-$ 11.75 gave reasonably accurate 
results\footnote{\citet{kirkpatrick2015} derived 
a similar template using a different technique, fitting the photometry 
of multiple galaxies over a range of redshifts.}
for all galaxies with infrared luminosities above $\sim$ 
10$^{11}$ L$_{\odot}$. This conclusion indicates that these templates are a reasonably 
accurate representation of the FIR SED of these galaxies. We start with this result to 
provide the fiducial SED against which we will study the evolution with increasing redshift toward a FIR SED resembling more closely that of Haro 11.

\begin{figure}
\epsscale{1.150}
\plotone{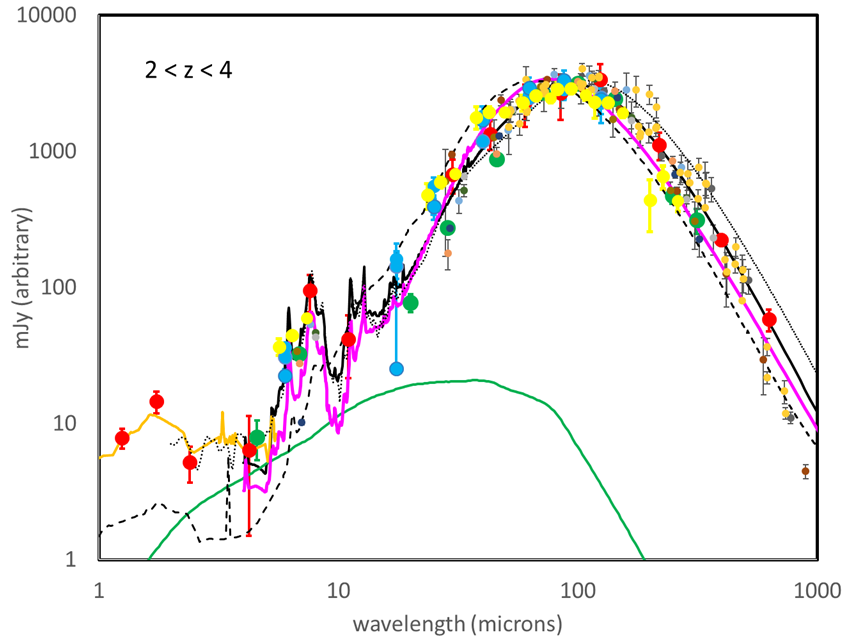}
\caption{Templates and data for infrared galaxies at $2 \le z < 4$. The solid black line is the template for log($L$(TIR)) = 11.25 and the 
purple line for log($L$(TIR))=11.75 from \citet{rieke2009}, and represent the range of these templates recommended at 
high redshift by \citet{rujopakarn2013}. The large red points are for the combined {\it Spitzer} + {\it Herschel} + ALMA 
photometry from \citet{dunlop2017}, median $z = 2$. 
They have been fitted at the shortest wavelengths by a stellar SED (orange) normalized to optimize the fit, and a 
``normal'' quasar SED (green)  \citep{lyu2017} has been subtracted. The large green points are from \citet{wang2016}, JH-red SF HIEROS; they have a median $z = 2.5$. The large yellow points are from \citet{bethermin2015} for stacked data in the COSMOS field. The large blue points are for the stacked results in the CANDELS fields from \citet{schreiber2018} for $2.5 < z < 3.5$; all three mass ranges are shown because the differences are minimal. The other (small) points are for individual galaxies, listed in Table 1. The dashed line is the template for $1 \times 10^{13}$ L$_\odot$ from \citet{chary2001} and the dotted one is the SFG2 template from \citet{kirkpatrick2015}. The template for $z = 3$ from \citet{schreiber2018} is not plotted for clarity; it falls very close to the \citet{rieke2009} log($L$(TIR)) = 11.25 one for $\lambda < 100 \mu$m and the log($L$(TIR)) = 12.25 one \citep{rieke2009} (not shown here), for $\lambda > 100 \mu$m. The template from \citet{magdis2012} is also not plotted; it falls close to the \citet{rieke2009} log($L$(TIR)) = 11.25 one.
\label{fig5}}
\end{figure}

In general, well-sampled far infrared SEDs are not available at very high 
redshifts, so our method is to collect all cases where there are 
measurements of reasonably high signal to noise covering a sufficient 
wavelength range to be useful in constraining the shape of the far infrared 
SED. We also require spectroscopic redshifts to avoid artificially broadening the SED due to redshift errors. 
Table 1 lists the galaxies and the sources for the measurements we have 
used. Most of the galaxies are selected at long wavelengths, although in the rest frame
there are many examples bridging the peak of the FIR SED. Specifically,  \citet{wang2016} select at 160~$\mu$m, 
\citet{casey2012} at 250 -- 500~$\mu$m, \citet{shu2016} at 500~$\mu$m, and the remaining 
cases in this redshift range are selected in the mm-range. 

We shift the measurements to the rest frame and normalize them to the 
same value (relative to the assumed template) near the peak. This normalization uses $\chi^2$ minimization 
to the template between rest wavelengths of 50 and 200~$\mu$m, i.e. $\chi^2$ is minimized using all the measurements 
that fall within this spectral range. To support this step, we require at least three measurements with signal to noise ratios $>$ 3 in this 
spectral range. Since this range covers the peak of the SED, the normalization is roughly by far infrared luminosity. The normalization depends only on the peak of the FIR SED, so it leaves the measurements at shorter and longer wavelengths to characterize the SED shape.  

Figure~\ref{fig5} shows the result for $2\le z<4$, along with the log($L$(TIR)) $=$ 11.25 and 11.75 templates \citep{rieke2009}, as well as templates from \citet{chary2001} and \citet{kirkpatrick2015}. Some of the points are from combinations of data for a number of galaxies. The composite 
data points from \citet{dunlop2017} are based on galaxies detected in a deep ALMA map of the Hubble Ultra-Deep Field. The 
red HIEROS from \citet{wang2016} are massive, dusty, star forming galaxies at redshifts of 2 -- 3 (plus 
passive galaxies at higher redshifts that will not contribute to the far 
infrared); the median redshift of these galaxies is $z=2.5$ and the quoted 
errors are about the same size as the points. \citet{bethermin2015} present stacked measurements in the COSMOS field; we show results from the bins between $z = 2$ and $z = 3.5$. \citet{schreiber2018} stacked {\it Herschel} measurements of galaxies 
in the CANDELS fields; we show the results for $2.5 < z <3.5$.  The remaining points are for individual galaxies. 

The values of $\chi^2$ provide a quantitative evaluation of the validity of the templates. A few measurements have very high ratios of nominal 
signal to noise, up to 20:1. To allow for effects such as confusion noise and at the same time
to keep one or two measurements from dominating the results, we added a uniform additional
noise of 15\% of the measured total flux density value for the individual galaxies and 7.5\% for the stacked results, in quadrature with the quoted noise. We also reject the 5 most discordant measurements (out of a total of 163), and evaluate the templates only between 11 and 700 $\mu$m. The log($L$(TIR)) = 11.25 template yields $\chi_{\rm red}^2$ = 1.87. The agreement is significantly worse with the log($L$(TIR)) = 11.75 template ($\chi_{\rm red}^2$ = 2.91) and with the Chary \& Elbaz  ($\chi_{\rm red}^2$ = 5.61), Kirkpatrick ($\chi_{\rm red}^2$ = 4.06), and Schreiber ($\chi_{\rm red}^2$ = 3.05) ones. The \citet{magdis2012} $z = 2.5$ template is also somewhat worse, $\chi_{\rm red}^2 = 2.27$. These values of $\chi_{\rm red}^2$ could be reduced by assigning a value larger than 15\% to the additional noise, but we have not done that because there are likely to be intrinsic variations among the galaxies relative to a single template, and they will contribute to the scatter. We will use the log($L$(TIR)) = 11.25 template as the standard of comparison in the following sections to demonstrate any systematic changes. However, it appears to overestimate the strength of the aromatic bands modestly, consistent with the finding of \citet{rujopakarn2013} that the rest $\sim$ 8 $\mu$m fluxes require templates with log($L$(TIR)) = 11.25 $-$ 11.75 for good fits. Since this paper is focused on  wavelengths $\ge$ 11 $\mu$m, this is not an issue.

\subsection{Far Infrared SEDs at Higher Redshift}

We now apply a similar approach to luminous infrared galaxies at $5 < 
z <7$. The measurements of individual galaxies with adequate far infrared observations are listed in 
Table 1. We also used measurements of six quasars with adequate signal to noise {\it Herschel} 
FIR measurements and a large excess attributed to their host 
galaxies, namely J0338$+$0021, J0756$+$4104, J0927$+$2001, J1202-3235, 
J1204-0021, and J1340$+$2813; flux densities and redshifts were taken from \citet{leipski2014}. 
We subtracted the quasar contribution using the ``normal'' quasar template 
(\citealt{xu2015}), as opposed to dust-deficient quasar templates (see \citealt{lyu2017}), to derive the SEDs of the host galaxies. 
None of the fits in \citet{lyu2016} would ascribe a significant fraction of the FIR emission 
to the quasar, a conclusion that is confirmed by arguments in Appendix B. 

\begin{figure}
\epsscale{1.15}
\plotone{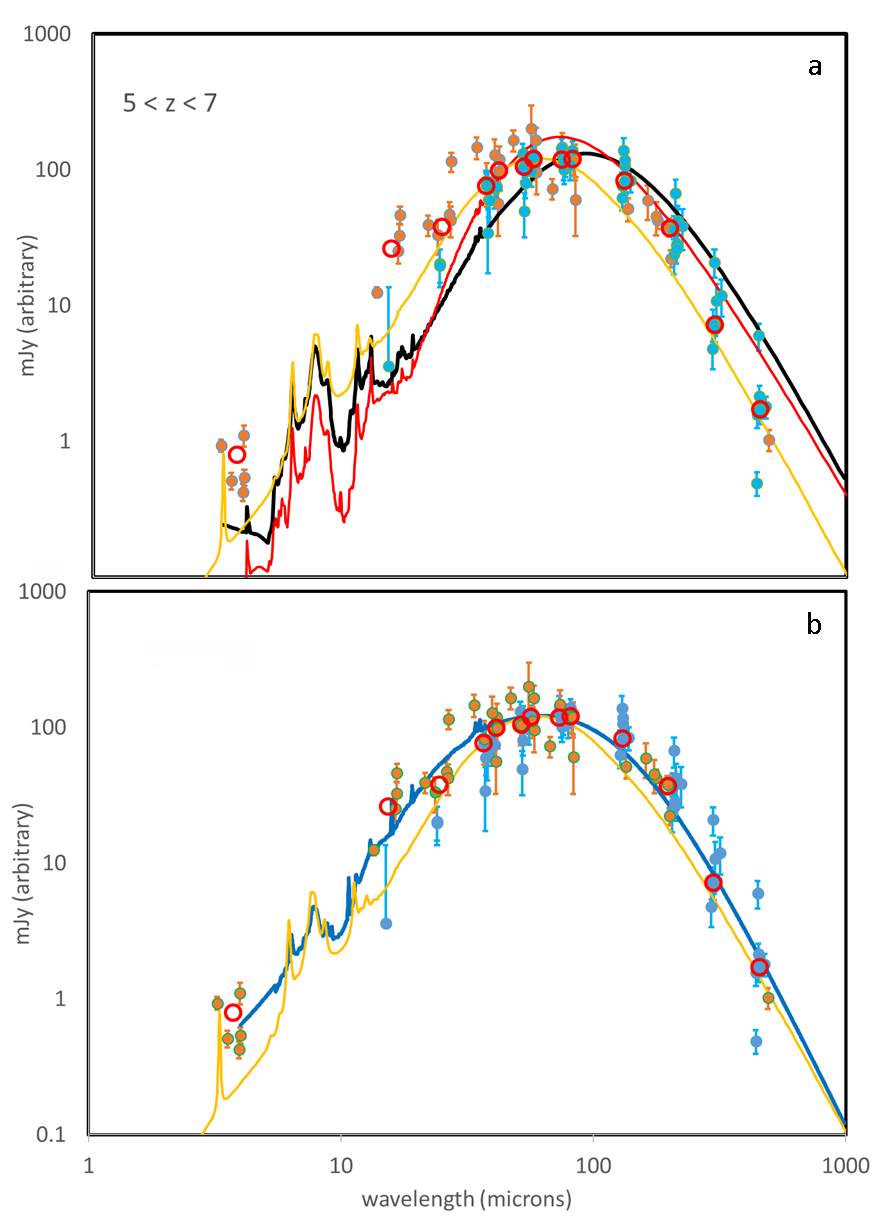}
\caption{(a) Measurements of galaxies at $5 < z <7$, shifted to the rest frame and compared with the log($L$(TIR)) = 11.25 (black) \& 12.25 (red) templates 
from \citet{rieke2009}, plus the $z = 6$ template from \citet{schreiber2018} (orange). The deep orange 
points are for the quasar host galaxies and the blue ones for galaxies not selected for AGN.  Medians in groups of 5 - 11 
measurements at adjacent wavelengths are shown as red circles. The $\chi_{\rm red}^2$ for the log($L$(TIR)) = 11.25 fit is 6.1, for log($L$(TIR)) = 12.25 it is 4.5, and for the Schreiber template it is 3.25 (see text for details). The points at $\lambda <$ 5.5~$\mu$m may have significant contributions by direct stellar emission and should not be considered as part of the far infrared SED.  The measurements suggest that these templates are all significantly too narrow at this redshift range.  (b)  Measurements as in (a), compared with the SED of Haro 11 (dark blue line). The $\chi_{\rm red}^2$ for the fit is 1.57 (see text for details), indicating a much better fit than in (a).
\label{fig6}}
\end{figure}

To look for changes in far infrared SEDs between $z \approx 3$ and $6$,  in Figure~\ref{fig6}(a)
we compare all of the available measurements with adequate signal to noise at the latter redshift 
to the local log($L$(TIR)) = 11.25 template that gave the best 
fit for $2 < z <4$. The correspondence is poor, largely because the SED is too cool. The poor fit would also apply 
to the approach of \citet{magdis2012}, whose template resembles the log($L$(TIR)) = 11.25 one, and who argue for 
little SED evolution at $z > 3$. 
The local log($L$(TIR)) = 12.25 template has a warmer SED, but nonetheless it has significant divergence at both short and long wavelengths. 
\citet{schreiber2018} extrapolate the temperature relation derived for $z < 4$ to deduce 
an SED at $z = 6$ with a temperature about 10\,K warmer than at $z = 3.5$. 
Figure ~\ref{fig6}(a) shows their template for $z = 6$; the correspondence with the data is improved but still not fully satisfactory.
To render these impressions quantitative, we follow the procedures in Section~4.1 to determine $\chi^2$ for the templates in  Figure~\ref{fig6}(a). We found a $\chi_{\rm red}^2$ of 6.1 for the log($L$(TIR)) = 11.25 template, 4.5 for the log($L$(TIR))=12.25 one, and 3.25 for the one from \citet{schreiber2018}. None of them fits well because the SEDs are too narrow. The measurements indicate the need for a  broader template. 

\begin{figure}
\epsscale{1.150}
\plotone{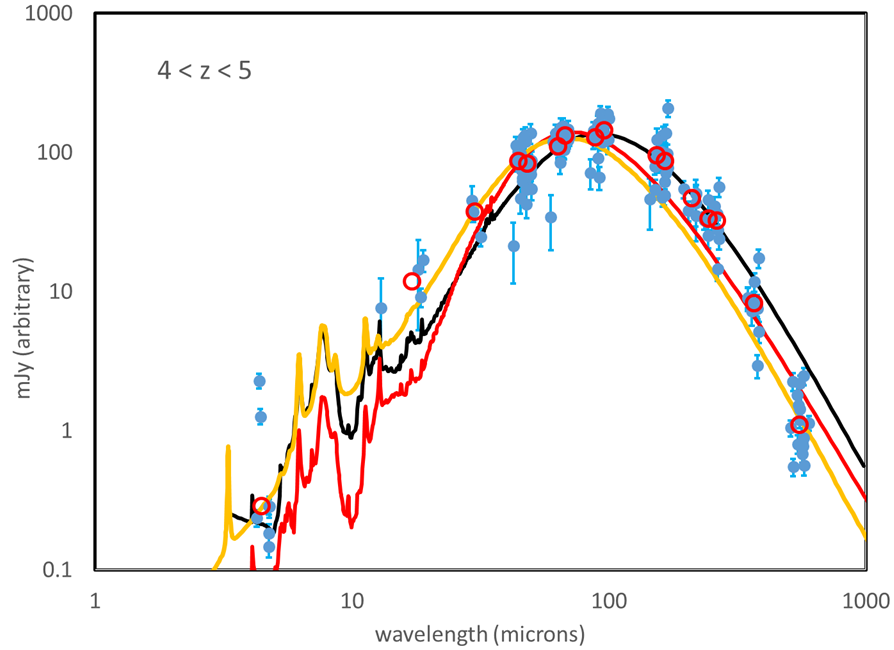}
\caption{Measurements of galaxies at $4 < z < 5$, shifted to the rest frame and compared with the log($L$(TIR)) = 11.25 (black) and 12.25 (red) templates 
from \citet{rieke2009} and with that from \citet{schreiber2018} for $z = 4.5$ (i.e., the average of those for $z = 4.4$ and 4.6). Medians of the measurements in groups of 3 - 12  measurements at adjacent wavelengths are shown as red circles.
\label{fig7}}
\end{figure}


Figure~\ref{fig7} is similar to Fig.~\ref{fig6}, but for  $4\le z <5$. It 
appears that the SEDs at  $z\approx $ 4.5 are a bridge from the local templates to the 
broader one needed to fit the data at $z > 5$. 
Figure \ref{fig6}(b) shows the measurements in Fig.~\ref{fig6}(a), fitted by the Haro 11 SED by $\chi^2$ minimization, using the same procedures as in Section~4.1.  The fit  has a $\chi_{\rm red}^2$ of 1.57.
There appears to be no systematic difference between the points derived from quasar host
galaxies and those from galaxies without luminous AGN. However, the points between 10 and
50~$\mu$m are mostly derived from the host galaxies because only they have useful measurements
in the 100 and 160~$\mu$m Herschel bands. Appendix B demonstrates that it is very unlikely
that these values are significantly contaminated by quasar emission. For comparison, we also show the relevant 
template from \citet{schreiber2018}. The Haro 11 SED is broader, and that results in a better fit to the measurements as 
indicated by its lower $\chi^2$, 1.57 vs. 3.25, when computed in identical ways. To convert the relative $\chi_{\rm red}^2$ values to probabilities, we 
increased the additional noise component from 15\% to 22\% of the measured flux, to set $\chi_{\rm red}^2$ for the Haro 11 template to 1. 
The probability to obtain by chance the resulting $\chi_{\rm red}^2$ for the Schreiber template, $\chi_{\rm red}^2$ = 2.10, is then  $\sim 10^{-8}$. 
This probability is reduced as the additional noise term is reduced from 22\% toward 15\%.

We conclude that the Haro 11 SED provides a significantly better template for galaxies at $z >5$, compared with templates derived at lower redshift.
Other than the quasar host galaxies, the galaxies in Fig.~\ref{fig6} are nearly 
all selected in the mm-waveband, corresponding to rest wavelengths around 150~$\mu$m, which could produce a mild 
bias toward ``cold'' SEDs. Nonetheless, their SEDs are significantly brighter in the mid-infrared ($10 - 40 \mu$m) and hence ``warmer" 
than is indicated by templates derived at lower redshift, as shown in Fig.~\ref{fig6}. 

The slope of the Haro 11 SED at wavelengths 
longer than 100 $\mu$m may be slightly steeper than those of other templates. To set an upper limit for the CMB contribution to the dust temperature, 
we estimated the emissivity index $\beta$ (where the departure from gray emissivity goes as 
$\lambda^{-\beta}$) just for wavelengths $\ge$ 100 $\mu$m, obtaining $\beta = 1.2$. A more
conventional estimate gives $\beta \approx 1.9$ \citep{lyu2016}, compared with values of 1.5 $-$ 2.0 for typical high metallicity main sequence 
luminous galaxies \citep[e.g.,][]{elbaz2011, magdis2012, dacunha2013}. Any differences in the sub-mm SEDs would be subtle, making it difficult to distinguish the two cases in that spectral range. Therefore, this potential bias does not undermine the 
conclusion that the Haro 11 template is the better template to use at these redshifts.

Our results show that the SEDs of infrared galaxies continue to evolve toward warmer temperature going from $z \sim 3$ to $z \sim 6$, in agreement with the extrapolation in \citet{schreiber2018}, but contrary to the assumptions in \citet{elbaz2011} and \citet{magdis2012}. The extent of the potential change in shape of the far infrared SEDs for $z \gtrsim 5$ is not anticipated in any previous sets of templates. 

\section{Implications for Luminosity Estimation}

\begin{figure}
\epsscale{1.150}

\plotone{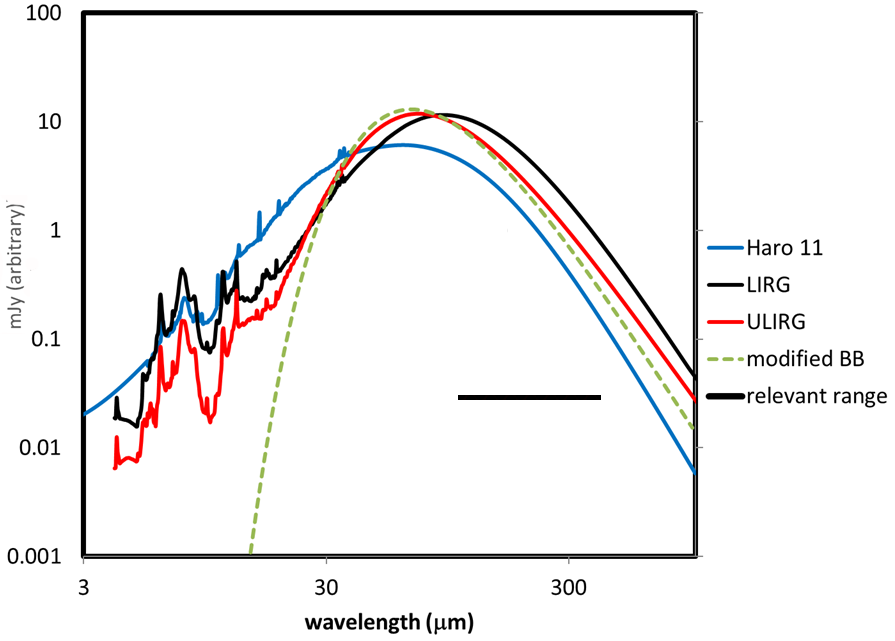}
\caption{Influence of template selection on SFR estimates from ALMA measurements at $\sim$ 1\,mm. All of the templates 
have been normalized to the same total infrared luminosity. The modified blackbody is as in \citet{wang2007, wang2008, leipski2014, willott2013, willott2017}. The horizontal black bar shows the most relevant wavelength range ($100- 400\,\mu$m rest) for interpreting single-band measurements near $\lambda_{\rm observed}\sim 1$\,mm of galaxies at $z\sim 1 - 10$. 
\label{fig9}}
\end{figure}

\begin{figure}
\epsscale{1.150}
\plotone{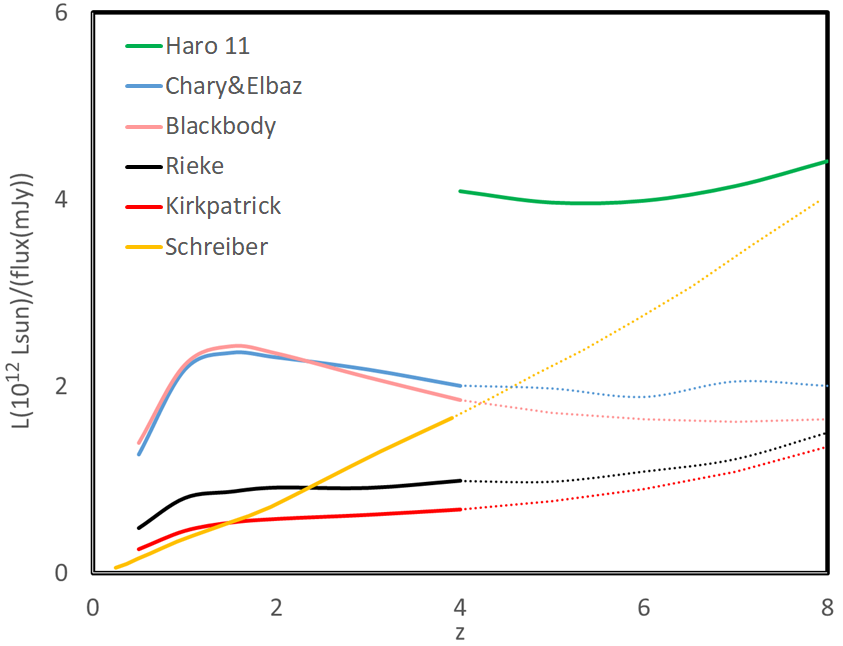}
\caption{Conversion factors from ALMA oberved Band 7 flux densities in mJy to L(TIR) in units of $10^{12}$ L$_\odot$, for various SED templates. The conversions follow the approach of \citet{schreiber2018}; given a flux density in mJy, if it is multiplied by the appropriate conversion factor for the redshift of the galaxy, the result is in units of $10^{12}$ L$_\odot$. The blackbody curve is for T = 47\,K, $\beta$ = 1.6, the Chary \& Elbaz template is for $10^{13}$ L$_\odot$, and the Kirkpatrick template is their SFG2 template. Beyond z = 4, all the templates except that based on Haro 11 are extrapolated, and hence are shown as dotted lines.
\label{fig10}}
\end{figure}

Spectral templates are central in determining average, or ``typical,'' properties of high-redshift galaxies. Perhaps the most critical use of them is determining total infrared luminosities, which for luminous, dust-shrouded galaxies provides an estimate of the luminosity in young and massive stars. Combined with the initial mass function, one can then obtain the star formation rate for a galaxy. Here we consider the impact of using the Haro 11 template on such estimates in two cases: (1) where Herschel measurements at 250 $\mu$m and longer wavelengths are available; and (2) when there is only a single-band measurement near 1 mm with ALMA. 

\subsection{Luminosities from Herschel measurements}

 The influence of template choice on the deduced far infrared luminosity with Herschel measurements is not uniquely determined; it depends on the 
bands measured and the method used to normalize the templates to the measurements (i.e., there is often an additional error term above the nominal measurement errors). For a specific example, however, we have assumed a galaxy at $z = 5, 6,$  or $7$ with measurements at 250, 350, and 500 $\mu$m (typical for such data on galaxies at $z\sim 6$) and equal weights. We then find that the luminosity deduced with the \citet{rieke2009} LIRG and ULIRG templates and the \citet{schreiber2018} one for $z = 5$ would all be about 77\% lower than with the Haro 11 one. At $z = 6$, the results would be 99, 68, and 77 \% of that from the Haro 11 template respectively. At $z = 7$ they are respectively 121, 79, and 82\%. These values are only illustrative, given the differences that would result from different measurement weights.

\subsection{Luminosities from single-band ALMA measurements}

Single-band ALMA measurements near 1 mm are being used to estimate infrared luminosities and star 
formation rates at $z \approx 6$ \citep[e.g.,][]{wang2013, willott2017, decarli2018}. 
One consequence of the far infrared SED derived here for high-redshift galaxies is that such an approach could systematically underestimate the total infrared luminosities and hence the SFRs. Figure \ref{fig9} compares a number of relevant SEDs to illustrate the issue. 
The LIRG spectrum (log(L(TIR))=11.25) has been shown to be typical of even the most luminous galaxies at $1<z<3$ (Rujopakarn et al. 2013, this paper). The 
modified black body (T$_{d} =$ 47\,K, $\beta =1.6$) is commonly 
used to interpret the ALMA 1mm measurements \citep[e.g.,][]{wang2007, wang2008, leipski2014, willott2013, willott2017}. The 
horizontal black line extends from 100 to 400\,$\mu $m, roughly the relevant 
rest spectral range for the interpretation of the ALMA bands 6 and 7 results at $z\sim 1-10$. The various possible choices for the SED result in significant 
differences in the total infrared luminosities that would be deduced.

This issue is illustrated graphically in Figure \ref{fig10}, which shows the conversion from flux densities measured in ALMA Band 7 (870\,$\mu$m) 
to galaxy luminosities in units of $10^{12}$ L$_\odot$, as in \citet{schreiber2018}. The differences in the conversion factors for objects at $1<z\le 4$ can be understood from Figure~\ref{fig5}, which compares the different templates with observed photometry and confirms that the \citet{rieke2009} template is a good overall fit to the available measurements at these redshifts. 
The \citet{kirkpatrick2015} template agrees closely with the \citet{rieke2009} template at wavelengths $<$ 100$\mu$m, which account for much of the total IR luminosity, but is significantly brighter at longer wavelengths. Consequently, a given luminosity yields a larger signal at the long wavelengths probed by ALMA Band 7 over this range of redshift, corresponding to observed wavelengths of 400 -- 170~$\mu$m. In other words, a given signal in mJy needs a smaller multiplication factor to be converted to the total infrared luminosity using the Kirkpatrick template compared with the Rieke one. In contrast, the template from \citet{chary2001} is higher at $\lambda<100\,\mu$m and lower at longer wavelengths compared to the \citet{rieke2009} template, such that a given signal in Band 7 requires a substantially higher total infrared luminosity. As shown in Figure~\ref{fig9}, the modified blackbody is also higher at $\lambda<100\,\mu$m and lower longward of this wavelength compared to the \citet{rieke2009} template, resulting in a similar behavior. The differences between the \citet{chary2001} and \citet{kirkpatrick2015} templates are a factor of $3-4$ over this redshift range. The \citet{schreiber2018} results are based on templates that evolve in shape with redshift, and hence the behavior differs from the rest. None of these templates are constrained by data at $z>4$, so we show the results at higher redshifts as dotted lines. 

At $z \sim 6$, the Haro 11 SED appears to be a better fit to the data than any of the templates derived at lower redshift, and total luminosities derived using it on the basis of ALMA Band 7 measurements will be a factor of $2 - 4$ higher than those from the extrapolations of the other templates. At redshifts of $z\sim 4-5$, it is possible that there is a mixture of behavior in the far infrared, making the total infrared luminosity estimation for a typical case ambiguous.  

\section{Conclusions}

We have modeled the far infrared spectral energy distribution (SED) expected from Population II galaxies undergoing the first phase of vigorous star formation after the Population III stage. For the first 300 $-$ 400 Myr, the interstellar dust in such systems will be silicate-rich, differing significantly from the dust in 
more mature galaxies.  We have then compared the models with observations of galaxies as a function of redshift. We find:

\begin{itemize}

\item{The modeled SEDs are shifted substantially to the bluer (``warmer'') wavelengths compared to the SEDs of local galaxies.}

\item{Although other factors contribute, e.g., the high energy density in the galaxy, this behavior arises in part because of the relatively high emission efficiency of silicates between 8 and 60~$\mu$m.}

\item{The SED of Haro 11 is similar to the theoretical prediction for silicate-rich dust with a small addition of carbon dust. Haro 11 is a local galaxy with a very young and vigorous starburst that probably reproduces the relevant conditions in the Population II galaxies.  }

\item{There is a progression with redshift in observed galaxy SEDs, from those resembling local ones at $2 \le z < 4$ to a closer resemblance to that of Haro 11 for $5\le z< 7$.}

\item{Estimates of total infrared luminosities at $z \gtrsim 5$ based on measurements at 250, 350, and 500 $\mu$m (i.e., with {\it Herschel} SPIRE) 
with templates that do not reflect the breadth of the Haro 11 SED may differ from those using it by $20 - 30$ \%.} 

\item{Estimates of the total infrared luminosities at $z\gtrsim 5$ (and hence star formation rates) with measurements near 1~mm in wavelength can vary by factors of 2 $-$ 3 or more depending on the SED template used. Currently popular modified blackbodies or local templates can result in significant underestimates compared with those using Haro 11 as the template.}

\end{itemize}

Further understanding of the far infrared SEDs at high redshift can come with measurements of hydrogen recombination lines to 
determine the populations of luminous stars independently of the far infrared and testing for consistency. 
These measurements will be enabled by {\it JWST}, which can 
measure both the Balmer and Paschen series lines. Use of secondary indicators of star formation, such as the radio-infrared relation, 
is problematic because their calibration may evolve with redshift. Direct measurements of the far infrared SEDs over 
the relevant wavelength region to probe Haro-11-like SEDs  ($100 - 300 \mu$m) will be challenging; the Space Infrared Telescope for Cosmology and Astrophysics (SPICA) will be helpful but in the end its contributions will be limited by confusion noise. 
The larger aperture of the Origins Space Telescope (OST) would 
allow extension of the understanding of these SEDs to significantly lower luminosities than is possible in other ways. More fully exploring the FIR frontier at the highest redshifts will crucially depend on the future availability of these next-generation facilities.

\section*{Acknowledgements}

We thank Rob Kennicutt, Dan Marrone, Karl Misselt, and Corentin Schreiber for helpful discussions.
MEDR acknowledges support from PICT-2015-3125 of ANPCyT.
VB acknowledges support from NSF grant AST-1413501. The work of GHR, IS, and JL was partially supported by NASA Grant NNX13AD82G, 
and that of IS was also partially supported by a Hubble Fellowship. 
We thank Alexander Ji for providing tabulated dust opacities for the dust model used here.
This work makes use of the Yggdrasil code \citep{zackrisson2011}, which adopts
Starburst99 SSP models, based on Padova-AGB tracks \citep{leitherer1999, vazquez2005}
for Population~II stars.

\eject


\eject
\begin{longtable}{lll}
\tabletypesize{\scriptsize}
\tablecaption{Galaxies with FIR SED Measurements
\label{obs}}
\tablehead{
\colhead{name} & \colhead{z} & \colhead{reference}
}
\startdata

GH500.13 & 1.99 & \citet{shu2016}\\
GH500.28 & 1.99 & \citet{shu2016} \\
GH500.12 & 2.00 & \citet{shu2016} \\
HUDF composite & $\sim$ 2 & \citet{dunlop2017} \\
GH500.16 & 2.02 & \citet{shu2016} \\
1HERMES X24 J095917.28+021300.4 & 2.101 &\citet{casey2012}  \\
GH500.21 & 2.13 & \citet{shu2016} \\
COSMOS  &  2.25  &  \citet{bethermin2015}  \\
SPT0002-52 &2.3523 & \citet{strandet2016} \\
GH500.33 & 2.41 &  \citet{shu2016} \\
GH500.14a & 2.47 &\citet{shu2016}  \\
GH500.26 &2.49 & \citet{shu2016} \\
Red HIEROS & $\sim$ 2.5 & \citet{wang2016} \\
GH500.9 & 2.57 &  \citet{shu2016}  \\
1HERMES X24 J095948.00+024140.7 & 2.60 & \citet{casey2012} \\
GH500.20 & 2.67 &  \citet{shu2016} \\
COSMOS  &  2.75  &  \citet{bethermin2015}  \\
SPT2349-50 & 2.877 & \citet{strandet2016} \\
CANDELS stack  &  3  &  \citet{schreiber2018}  \\
SDP81 & 3.04 &  \citet{negrello2010}, \citet{yuan2015}  \\
SPT2357-51 & 3.0703 & \citet{strandet2016} \\
Eye & 3.074 & \citet{saintonge2013}, \citet{yuan2015}\\
SPT0103-45 & 3.0917 & \citet{weiss2013}, \citet{yuan2015} \\
SPT2307-50 & 3.108 &  \citet{strandet2016} \\
1HERMES X24 J033136.96−275510.9 & 3.145 & \citet{casey2012}  \\
SMMJ16354+66114 & 3.188  & \citet{magnelli2012}, \citet{yuan2015}  \\
A68/nn4 & 3.19 &  \citet{sklias2014}, \citet{yuan2015} \\
COSMOS  &  3.25  &  \citet{bethermin2015}  \\
1HERMES X24 J160603.63+541245.1 & 3.331 &  \citet{casey2012}   \\
SPT0529-54 &3.3689 & \citet{weiss2013}, \citet{yuan2015}  \\
1HERMES X1.4 J123536.28+623019.9 & 3.38 &  \citet{casey2012}   \\
SPT0532-50 & 3.3988 & \citet{weiss2013}, \citet{yuan2015}  \\
1HERMES X24 J160802.63+542638.1 & 3.415 &  \citet{casey2012}  \\
1HERMES X24 J160539.72+534450.3 & 3.546 &  \citet{casey2012}   \\
1HERMES X1.4 J100024.00+021210.9 & 3.553 &  \citet{casey2012}   \\
1HERMES X1.4 J123622.58+620340.3 & 3.579 &  \citet{casey2012}  \\
SPT2147-50 & 3.7602 & \citet{weiss2013}, \citet{yuan2015}  \\
HELMS-RED-31 & 3.798 & \citet{asboth2016}, \citet{oteo2017}\\
1HERMES X24 J160639.40+533558.4 & 3.801 &  \citet{casey2012}  \\
NGHP-246114 & 3.847 &  \citet{fudamoto2017b} \\
SPT2340-59 & 3.864 & \citet{strandet2016} \\
1HERMES X1.4 J100111.52+022841.3 & 3.975 &  \citet{casey2012}, \citet{yuan2015} \\
SGP-354388 & 4.002 & \citet{fudamoto2017b}, \citet{oteo2018} \\
GH500.18 & 4.04 & \citet{shu2016}  \\
GN500.27a & 4.04 &   \citet{shu2016} \\
GN10 & 4.05 & \citet{huang2014} \\
GN20 & 4.05 &  \citet{huang2014} \\
GN20.2 & 4.05 &  \citet{huang2014}  \\
G1 & 4.05 &\citet{roseboom2012}, \citet{yuan2015}\\
SPT0418-47 & 4.2248 &  \citet{weiss2013}, \citet{yuan2015}  \\
SPT0113-46 & 4.2328 &  \citet{weiss2013}, \citet{yuan2015}   \\
HATLASID141 & 4.24 & \citet{cox2011}, \citet{yuan2015}\\
SGP-261206 & 4.242 & \citet{fudamoto2017b},  \citet{oteo2017} \\
SPT2311-54 & 4.2795 & \citet{strandet2016} \\
SPT0345-47 & 4.2958 & \citet{weiss2013}, \citet{yuan2015}  \\
SPT2349-56 & 4.304 &  \citet{strandet2016} \\
Blue HIEROS & $\sim$ 4.4 & \citet{wang2016} \\
NGP-190387 & 4.42 & \citet{fudamoto2017b} \\
SGP-196076 & 4.425 & \citet{fudamoto2017b} \\
SPT2103-60 & 4.4357 &  \citet{weiss2013}, \citet{yuan2015} \\
X24 J095916.08+021215.3 & 4.454 &  \citet{casey2012}, \citet{yuan2015} \\
SPT0441-46 & 4.4771 &  \citet{weiss2013}, \citet{yuan2015}  \\
1HERMES X1.4 J104722.56+590111.7 & 4.521 & \citet{casey2012}, \citet{yuan2015} \\
G09-81106 & 4.53 &  \citet{fudamoto2017b} \\
SPT2146-55 & 4.5672 & \citet{weiss2013}, \citet{yuan2015} \\
Vd-17871 & 4.622 &  \citet{smolcic2015}, \citet{yuan2015}  \\
AzTEC1 & 4.64 & \citet{huang2014} \\
1HERMES X1.4 J104649.92+590039.6 & 4.71 & \citet{casey2012} \\
SPT2335-53 & 4.757 & \citet{strandet2016}  \\
SPT2132-58 & 4.7677 & \citet{weiss2013}, \citet{yuan2015} \\
NGP-284357 & 4.894 & \citet{fudamoto2017b}  \\
1HERMES X24 J161506.65+543846.9 & 4.952 & \citet{casey2012} \\
J0338+0021 & 5.00 & \citet{lyu2016} \\
J1204-0021 & 5.03 & \citet{lyu2016}  \\
J0756+4104 & 5.09 & \citet{lyu2016} \\
HeLMS-RED-4 & 5.162 & \citet{asboth2016}\\
HLSA773 & 5.24 & \citet{combes2012},  \citet{yuan2015}  \\
SPT2319-55  &  5.2929  &  \citet{strandet2016}  \\
AzTEC3 & 5.30 & \citet{huang2014}  \\
J1202+3235 & 5.31 & \citet{lyu2016} \\
J1340+2813 & 5.34 & \citet{lyu2016} \\
ADFS-27 &5.655 & \citet{riechers2017} \\
SPT0346-52 & 5.6559 & \citet{weiss2013}, \citet{yuan2015}, \citet{ma2015,ma2016} \\
SPT0243-49 & 5.699 & \citet{weiss2013}, \citet{yuan2015} \\
SPT2353-50  &  5.576  &   \citet{strandet2016}  \\
J0927+2001 & 5.77 &  \citet{lyu2016}  \\
SPT0459-59 & 5.7993 & \citet{weiss2013}, \citet{yuan2015} \\
SPT2351-57  &  5.811  &   \citet{strandet2016}  \\
G09-83808 & 6.027 & \citet{fudamoto2017b}, \citet{zavala2018}  \\
HFLS3 & 6.34 & \citet{riechers2014},  \citet{yuan2015} \\
\end{longtable}

\eject

\begin{appendices}

\section{Spectral energy distribution of Haro 11}

We provide a digital form of the infrared spectral energy distribution of Haro 11 in Table~\ref{harosed}. The first few entries are shown here for illustrative purposes.

\begin{deluxetable}{cc}

\tabletypesize{\tiny}
\tablecaption{Spectral energy distribution of Haro 11
\label{harosed}}
\tablehead{
\colhead{ \hspace{4cm} wavelength ($\mu$m) ~~~~~~~~~~~~~~} & \colhead{Jy} } 
\startdata
3.00262	&	2.04E-02	\\
3.00954	&	2.04E-02	\\
3.01648	&	2.05E-02	\\
3.02343	&	2.06E-02	\\
3.03040	&	2.06E-02	\\
3.03739	&	2.07E-02	\\
3.04439	&	2.07E-02	\\
3.05141	&	2.08E-02	\\
\enddata
\end{deluxetable}


\section{Potential AGN Contributions to the FIR SED}

In this appendix, we address whether the points between 10 and 50~$\mu$m 
derived by fitting high redshift quasar observations could give an 
overestimated measure of the galaxy output in this spectral range. \citet{lyu2016} tested fitting the far infrared SEDs of high redshift quasars 
with a combination of a FIR-star-formation-removed \citet{elvis1994}-like quasar continuum template and 
\citet{rieke2009} star forming templates for log(L(TIR)) $=$ 11.25 and 
12.50 and found that both types of fit fell far short of the measurements in 
the 10 -- 50~$\mu $m range. In contrast, the Haro 11 SED gave a good fit. If 
the ``local'' templates are used to fit the measurements of these quasars, 
then the quasar must be far brighter than the Elvis et al. SED in this 
spectral range to account for the output in the mid-infrared. 

All of the high-redshift quasars are of very luminous type-1 AGNs without evidence of significant reddening (the strong UV continuum of 
this type is essential to their discovery). \citet{lyu2017,lyu2017b} 
evaluated the AGN infrared SEDs of such quasars. In the first reference they found two 
general cases where the mid-infrared continua can have significantly less 
output in the 10 -- 20~$\mu $m range than the Elvis template and they showed 
that this behavior was independent of redshift up through z $\sim$
6. They dubbed these cases ``warm dust deficient (WDD)'', meaning a typical 
Elvis-like peak near 3~$\mu $m but a SED that drops more rapidly than the 
Elvis one toward longer wavelengths; and ``hot dust deficient (HDD)'', which 
does not have the 3~$\mu$m feature but simply drops relative to the Elvis 
template from there to longer wavelengths. If we have used the Elvis 
template where one of these alternatives would be appropriate, then we will 
have underestimated the potential star-forming contribution in the 10 -- 
50~$\mu $m range. 

\subsection{Association of the Warm Dust with Star Formation}

Nonetheless, it is worth considering in more detail the possibility that the 
warm far infrared spectral component that produces the deviation from 
``local'' templates in the 10 -- 50~$\mu $m range might be dust heated by the 
quasar. \citet{lyu2017b} advanced a number of arguments that show that 
the intrinsic SEDs of Type 1 quasars drop at wavelengths longer than 20 -- 30~$\mu $m. 
Some local AGNs have an additional emission component due to polar dust that 
boosts the emission in this region (e.g., \citealt{asmus2016}, \citealt{lyu2018}), but this 
component is largely missing in more luminous AGN, probably because of the 
larger radiation pressure in their polar directions that tends to eject 
material rapidly. Hypothesizing that this warm component is present in all 
six of the high redshift quasars used to derive the stellar-powered SED in 
this paper requires that their SEDs differ significantly from those of 
lower-redshift quasars, which would be contrary to all the evidence that the 
SEDs are identical, including in the infrared \citep[e.g.,][]{fan2009, lyu2017}. 

\begin{figure}
\epsscale{1.150}
\plotone{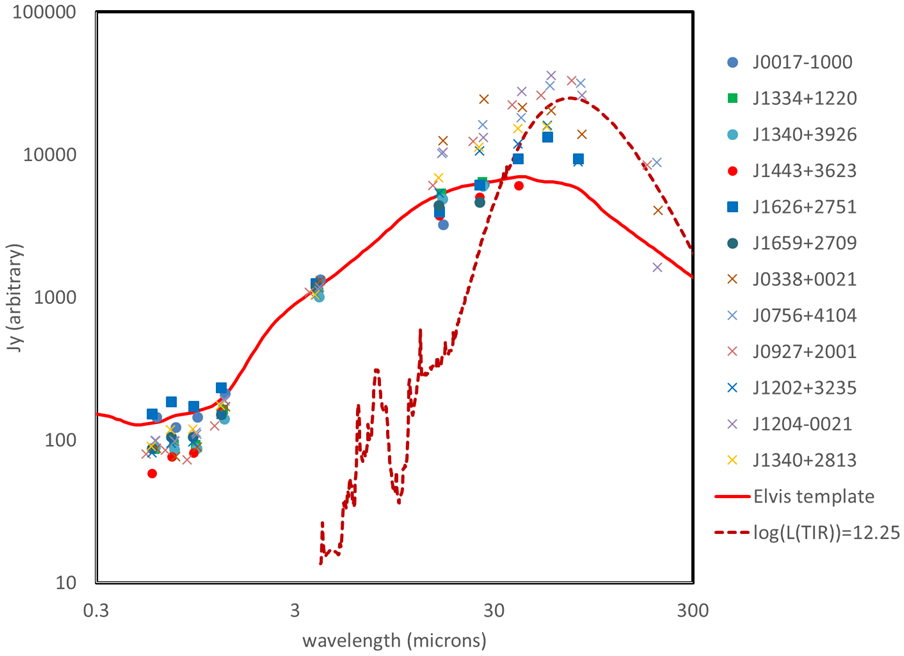}
\caption{Photometry of all high-z quasars in Leipski et al. (2014) with usable signal to noise ($>$ 3) in the far infrared. The measurements have been shifted to the rest frame and normalized near 4~$\mu$m. An Elvis SED has been normalized to the measurements at this wavelength and a “local” star forming template with log(L(TIR)) = 12.25 is shown roughly fitted to the points near the peak of this template. The galaxies with the largest far infrared excesses above the quasar continuum, lying around the peak of this “local” template (x symbols), are also the ones with significant excesses above the Elvis SED in the 10 $-$ 50~$\mu$m range. 
\label{fig11}}
\end{figure}

Figure \ref{fig11} illustrates the issue specifically for these six quasars. It shows 
that the strongest excesses above the quasar continuum template in the 10 -- 50~$\mu $m range are associated with powerful excesses at the longer far infrared wavelengths. These latter excesses  
almost certainly arise from dust heated by star formation in the host 
galaxies. This relationship is not expected if the quasar itself is 
responsible for the 10 -- 50~$\mu $m emission, in which case it would be 
expected to be independent of the output of the host galaxies.

\subsection{Energy Balance and the Warm Dust}

The hypothesis that this emission arises from heating by the quasar can also 
be tested for these specific objects by energy balance: is the expected 
optical to UV luminosity of the quasar sufficiently larger than the infrared 
luminosity to make it plausible that the infrared arises through heating of 
dust by the quasar? In general, the value of the ratio of these quantities 
should reveal the covering fraction of dusty structures around the AGN 
central engine. We have applied this test to the six quasars in this study. 
We used the log(L(TIR)) $=$ 12.25 template to represent the star formation in 
the host galaxies, since it provides a rough fit at the long wavelengths. To 
it, we added the output of a single temperature grey-body with a 
wavelength-dependent emissivity proportional to $\lambda^{\mathrm{-\beta 
}}$, with $\beta =$ 2. This fit should produce a lower limit to the 
luminosity of the warm component since it produces a SED peaked relatively 
sharply at the critical wavelengths; addition of components over a range of 
temperatures would increase the derived luminosity, for example, since they 
would add emission at wavelengths not well constrained by the data. We 
computed the energy balance assuming that all the luminosity at wavelengths 
short of 1.1~$\mu $m was available to heat the warm dust and that all the 
luminosity at longer wavelengths was from such heated dust except for the 
contribution from the ULIRG template. The rationale for this approach can be 
found in \citet{lyu2017b}. The results are in Table~\ref{balance}, shown as lower 
limits because of the potential underestimate of the warm component 
luminosity due to the assumption of a single temperature. 

With the simple assumption of isotropic emission, the first three quasars 
listed are already in violation of the expectations from energy balance. A 
more rigorous comparison needs to take account of the expected anisotropies 
in the emission by the central engine of an AGN. This case has been modeled 
by \citet{stalevski2016}, who studied the relation between the covering 
fraction and the reradiated accretion disk emission with a torus that is 
optically thick in the mid-IR. Their simulations indicate how the luminosity 
changes with different parameter values for the torus. For the cases where 
the predicted SED matches that observed, the upper limit to the ratio of 
reradiated to central engine luminosity is $\sim $ 0.75 (see the further 
discussion in \citealt{lyu2017b}). All six quasars are above this upper limit (see Table 2). 

The quasars in Table~\ref{balance} constitute about half of those in Leipski et al. 
(2014) with useful far infrared measurements, i.e. detections at 3:1 or more 
signal to noise in multiple bands. It is unlikely from an energy balance 
perspective that half of the quasars with useful Herschel data both have 
some unique circum-nuclear structure to emit in the mid-IR and are also 
pushing the limits of energy balance. Of course the energy balance argument 
could be circumvented if the ultraviolet spectra of these quasars are 
anomalously bright, but then we need to invoke two departures from typical 
quasar behavior (UV and mid-IR), when all the evidence to date is that the 
high-redshift quasars are identical in all observables to their lower 
redshift counterparts \citep[e.g.,][]{fan2009}.

\begin{deluxetable}{ccc}
\tabletypesize{\tiny}
\tablecaption{Energy balance between UV heating and possible warm dust component
\label{balance}}
\tablehead{
\colhead{\hspace{2cm}quasar ~~~~~~~~~~~~~~~~~~~~~} & \colhead{warm dust} & \colhead{IR reradiated luminosity/} \\
\colhead{} & \colhead{temperature} & \colhead{UV heating luminosity}
}
\startdata
J0338+0021  &  100  &  $>$ 1.15  \\
J0756+4104  &  120  &  $>$ 1.23  \\
J0927+2001  &  120  &  $>$ 1.32  \\
J1202-3235   &  125  &  $>$ 0.88  \\
J1204-0021  &  80  &  $>$ 0.86  \\
J1340+2813 &  110  &  $>$ 0.85  \\
\enddata
\end{deluxetable}

\end{appendices}

\end{document}